\documentclass[prc, reprint, amsmath, amssymb, superscriptaddress, showpacs]{revtex4-1}
\usepackage{natbib}
\usepackage{verbatim}
\usepackage{graphicx}
\usepackage{bm}
\usepackage{lipsum}
\usepackage{multirow}

\newcommand{\nuc}[2]{${}^{#1}\text{#2}$}
\newcommand{\unit}[1]{\ensuremath{\, \mathrm{#1}}}
\begin{document}

\bibliographystyle{apsrev}

\title{Precision Mass Measurements of Neutron-Rich Co Isotopes Beyond N=40}%

\author{C. Izzo}%
\email[]{izzo@nscl.msu.edu}
\affiliation{Department of Physics and Astronomy, Michigan State University, East Lansing, Michigan 48824, USA}
\affiliation{National Superconducting Cyclotron Laboratory, East Lansing, Michigan 48824, USA}

\author{G. Bollen}%
\affiliation{Department of Physics and Astronomy, Michigan State University, East Lansing, Michigan 48824, USA}
\affiliation{Facility for Rare Isotope Beams, East Lansing, Michigan 48824, USA}%

\author{M. Brodeur}%
\affiliation{Department of Physics, University of Notre Dame, Notre Dame, Indiana 46556, USA}

\author{M. Eibach}%
\affiliation{National Superconducting Cyclotron Laboratory, East Lansing, Michigan 48824, USA}
\affiliation{Institut f\"ur Physik, Universit\"at Greifswald, 17487 Greifswald, Germany}

\author{K. Gulyuz}%
\affiliation{National Superconducting Cyclotron Laboratory, East Lansing, Michigan 48824, USA}

\author{J. D. Holt}%
\affiliation{TRIUMF 4004 Wesbrook Mall, Vancouver, British Columbia V6T 2A3, Canada}

\author{J. M. Kelly}%
\affiliation{Department of Physics, University of Notre Dame, Notre Dame, Indiana 46556, USA}

\author{M. Redshaw}%
\affiliation{National Superconducting Cyclotron Laboratory, East Lansing, Michigan 48824, USA}
\affiliation{Department of Physics, Central Michigan University, Mount Pleasant, Michigan 48859, USA}
\affiliation{Science of Advanced Materials Program, Central Michigan University, Mount Pleasant, Michigan 48859, USA}

\author{R. Ringle}%
\affiliation{National Superconducting Cyclotron Laboratory, East Lansing, Michigan 48824, USA}

\author{R. Sandler}%
\affiliation{Department of Physics and Astronomy, Michigan State University, East Lansing, Michigan 48824, USA}
\affiliation{National Superconducting Cyclotron Laboratory, East Lansing, Michigan 48824, USA}
\affiliation{Department of Physics, Central Michigan University, Mount Pleasant, Michigan 48859, USA}
\affiliation{Science of Advanced Materials Program, Central Michigan University, Mount Pleasant, Michigan 48859, USA}

\author{S. Schwarz}%
\affiliation{National Superconducting Cyclotron Laboratory, East Lansing, Michigan 48824, USA}

\author{S. R. Stroberg}%
\altaffiliation[Current address: ]{Physics Department, Reed College, Portland, OR 97202, USA}
\affiliation{TRIUMF 4004 Wesbrook Mall, Vancouver, British Columbia V6T 2A3, Canada}

\author{C. S. Sumithrarachchi}%
\affiliation{National Superconducting Cyclotron Laboratory, East Lansing, Michigan 48824, USA}

\author{A. A. Valverde}%
\altaffiliation[Current address: ]{Department of Physics, University of Notre Dame, Notre Dame, Indiana 46556, USA}
\affiliation{Department of Physics and Astronomy, Michigan State University, East Lansing, Michigan 48824, USA}
\affiliation{National Superconducting Cyclotron Laboratory, East Lansing, Michigan 48824, USA}

\author{A. C. C. Villari}%
\affiliation{Facility for Rare Isotope Beams, East Lansing, Michigan 48824, USA}

\date{\today}%

\begin{abstract}
The region near $Z$=28, $N$=40 is a subject of great interest for nuclear structure studies due to spectroscopic signatures in \nuc{68}{Ni} suggesting a subshell closure at $N$=40.  Trends in nuclear masses and their derivatives provide a complementary approach to shell structure investigations via separation energies. Penning trap mass spectrometry has provided precise measurements for a number of nuclei in this region, however a complete picture of the mass surfaces has so far been limited by the large uncertainty remaining for nuclei with $N>$ 40 along the iron and cobalt chains. Here we present the first Penning trap measurements of \nuc{68,69}{Co}, performed at the Low-Energy Beam and Ion Trap facility at the National Superconducting Cyclotron Laboratory.  In addition, we perform ab initio calculations of ground state and two-neutron separation energies of cobalt isotopes with the valence-space in-medium similarity renormalization group approach based on a particular set of two- and three-nucleon forces which predict saturation in infinite matter. We discuss the importance of these measurements and calculations for understanding the evolution of nuclear structure near \nuc{68}{Ni}.
\end{abstract}

\maketitle

\section{Introduction}
One of the primary concerns of present nuclear structure research is the evolution of the nuclear shell model in regions of the nuclear chart far from stability.  In these regions, effects such as the tensor force \cite{Otsuka2005} and three-body interactions \cite{Zuker2003, Otsuka2010,Hebe15ARNPS} drive shifts in the relative spacing of shell model energy levels.  This has been seen to result in the disappearance of some of the nuclear magic numbers observed in stable nuclei and the appearance of new ones.  When smaller energy gaps arise within a nuclear shell (known as subshell gaps), an interesting and complex balance of midshell and closed shell effects may arise.

\nuc{68}{Ni} has elicited a great deal of discussion relating to the possible arrival of a new magic number at $N$=40.  In a pure harmonic-oscillator potential 40 is a magic number, but the more realistic nuclear case including spin-orbit interaction and using a Woods-Saxon potential shape brings the lowest level of the next harmonic oscillator shell, the $g_{9/2}$, down in energy close to the $pf$ shell, which leads to the well-established nuclear magic number 50 instead of 40.  The spacing between the $g_{9/2}$ orbital and the $pf$ orbitals is then the key question in examining possible subshell behavior at $N$=40.  In the proton case, the level structure of the \nuc{90}{Zr} nucleus supports a subshell closure at the proton number $Z=40$ \cite{Broda1995}.  Several experimental studies of neutron-rich Ni isotopes support the idea of a substantial neutron subshell closure at $N$=40 as well, including transfer reaction cross sections \cite{Bernas1982}, first excited $2^+$ state energies \cite{Broda1995}, and $B(E2)$ values \cite{Sorlin2002}.  However, this behavior disappears quickly moving away from Ni.

In addition to these reaction and spectroscopy studies, neutron shell closures and subshell closures may be studied using two-neutron separation energies ($S_{2n}$), defined as simply the difference between binding energies of a given nucleus and the nucleus with two fewer neutrons.  As nuclear binding energies account for the difference between the mass of a nucleus and the sum of its constituent nucleon masses, two-neutron separation energies can be expressed as a function of atomic masses:

\begin{equation}
S_{2n}(N,Z) = [m(N-2,Z) - m(N,Z) + 2m_n]c^2,
\end{equation}
where $m_n$ is the mass of a free neutron and $c$ is the speed of light.  The Weizs{\"a}cker mass formula \cite{Weizsacker1935} suggests a linear decrease in $S_{2n}$ as a function of neutron number $N$ for a given isotopic chain; deviations from this linear trend indicate additional structure effects \cite{Cakirli2009}.  Thus, precise measurements of atomic masses can be used for detailed nuclear structure studies.  In the particular case of a shell gap or subshell gap, a sudden drop in the two-neutron separation energies relative to the linear decline can be observed at the magic number.

With the use of on-line Penning trap mass spectrometry, separation energies have been determined to a high degree of precision (uncertainties $\sim$10~keV or better) for many nuclides in the \nuc{68}{Ni} region, extending up to $N$ = 40 in the $_{26}$Fe and $_{27}$Co chains and beyond $N$=40 in the $_{28}$Ni, $_{29}$Cu, $_{30}$Zn, and $_{31}$Ga chains \cite{Guenaut2007, Rahaman2007, Ferrer2010}.  These results demonstrate no subshell behavior across $N$=40 for $Z>$ 28.  In the Ni chain, the results are somewhat surprising, showing an unexpectedly large $S_{2n}$ value for $N$=39 and $S_{2n}$ values falling linearly across $N$=40, as shown in Fig.~\ref{fig4}.  Furthermore, the work of Grawe and Lewitowicz \cite{Grawe2001} demonstrates that the experimental spectroscopic data in \nuc{68}{Ni} often attributed to an $N$=40 subshell gap may be explained by the change in parity from the $pf$ shell to the $g_{9/2}$ orbital even with no energy gap.  

In order to develop a full picture of the structure in the region, measurements of the separation energies across $N$=40 for $Z<$ 28 are necessary.  However, up to now Penning trap measurements in this region have only extended up to $N$=40, leaving uncertainties larger than 100~keV in the separation energies beyond $N$=40 \cite{Wang2016}.  These large uncertainties obscure the trends at the critical point of crossing the $N$=40 subshell closure.  Here we report the first Penning trap measurements of $_{27}^{68}$Co and $_{27}^{69}$Co, reducing the mass uncertainties by more than an order of magnitude.

\section{Experiment}
\begin{figure}
\includegraphics[scale=0.35]{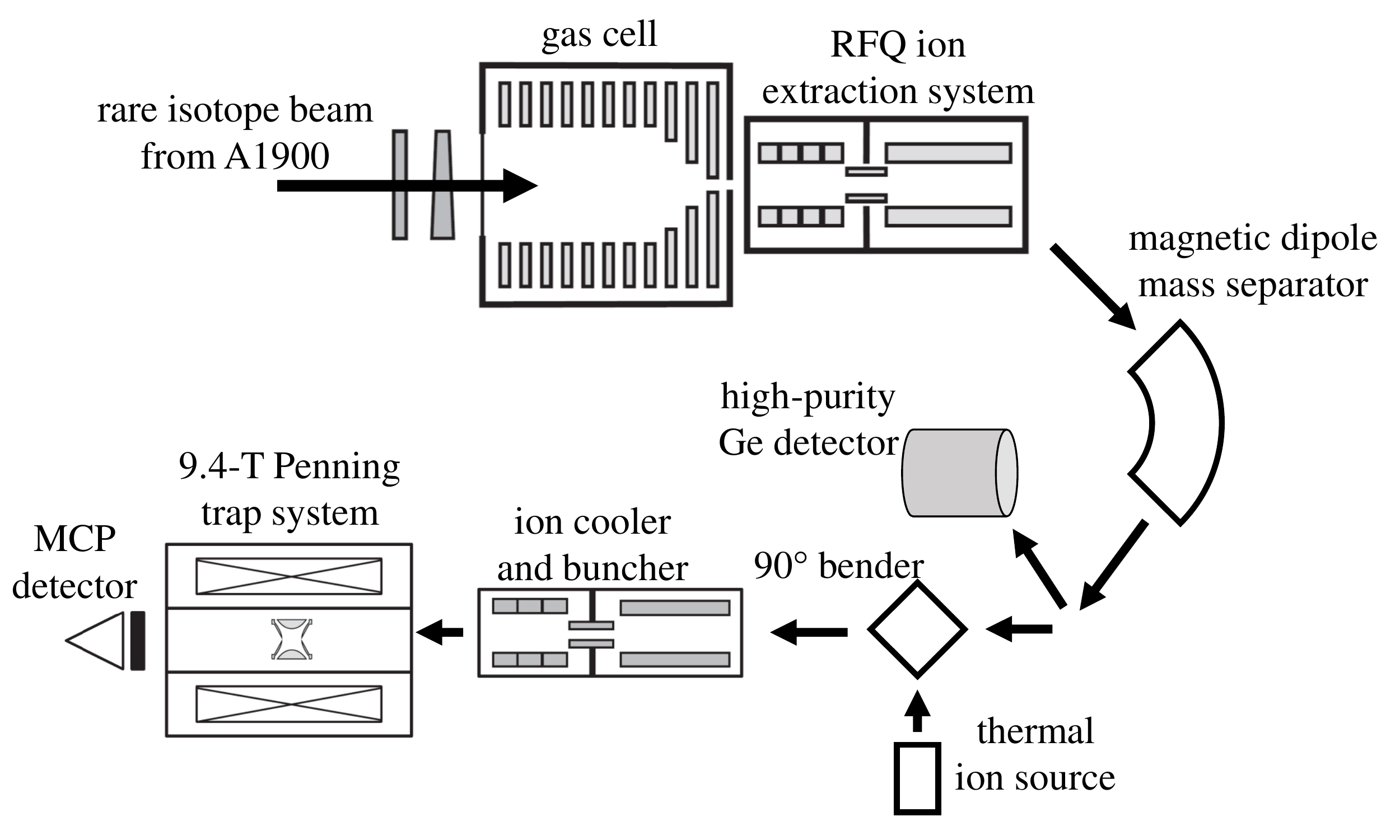}
\caption{\label{fig1}Schematic overview of the major experimental components used for this work.}
\end{figure}

The experiment was carried out at the National Superconducting Cyclotron Laboratory at Michigan State University.  A schematic view of the major experimental stages is presented in Fig.~\ref{fig1}.  \nuc{68}{Co} and \nuc{69}{Co} were produced by projectile fragmentation of a 130~MeV/u \nuc{76}{Ge} primary beam on a Be target.  The desired fragments were selected using the A1900 fragment separator \cite{Morrissey2003} and then slowed to an energy of less than 1~MeV/u by passing through a system of solid degraders before entering a gas cell filled with high-purity helium gas \cite{Cooper2014}.  Inside the gas cell, the ions were slowed to thermal energies by collisions with helium atoms.  A combination of gas flow and a DC electric field were used to transport the ions through the gas cell, while an RF electric field was used to repel the ions from the walls.  The ions were then extracted in a doubly-charged state through a radiofrequency quadrupole (RFQ) ion guide and passed through a magnetic dipole mass separator with a resolving power of $\frac{m}{\Delta m} \approx 1500$.  Doubly-charged Co ions were then sent to the Low-Energy Beam and Ion Trap (LEBIT) facility \cite{Ringle2013}.

The first major component of the LEBIT facility is a two-stage RFQ cooler/buncher \cite{Schwarz2016}, which takes the continuous beam from the gas cell and delivers short, low-emittance ion pulses to the 9.4-T Penning trap system.  A quadrupolar RF electric field is then applied to excite the ions' motion at a frequency $\nu_\text{RF}$ near the cyclotron frequency

\begin{equation}
\nu_c = \frac{qB}{2\pi m},
\end{equation}
where $q$ and $m$ are the charge and mass of the ion, respectively, and $B$ is the strength of the Penning trap magnetic field.

\begin{figure}
\includegraphics[scale=0.35]{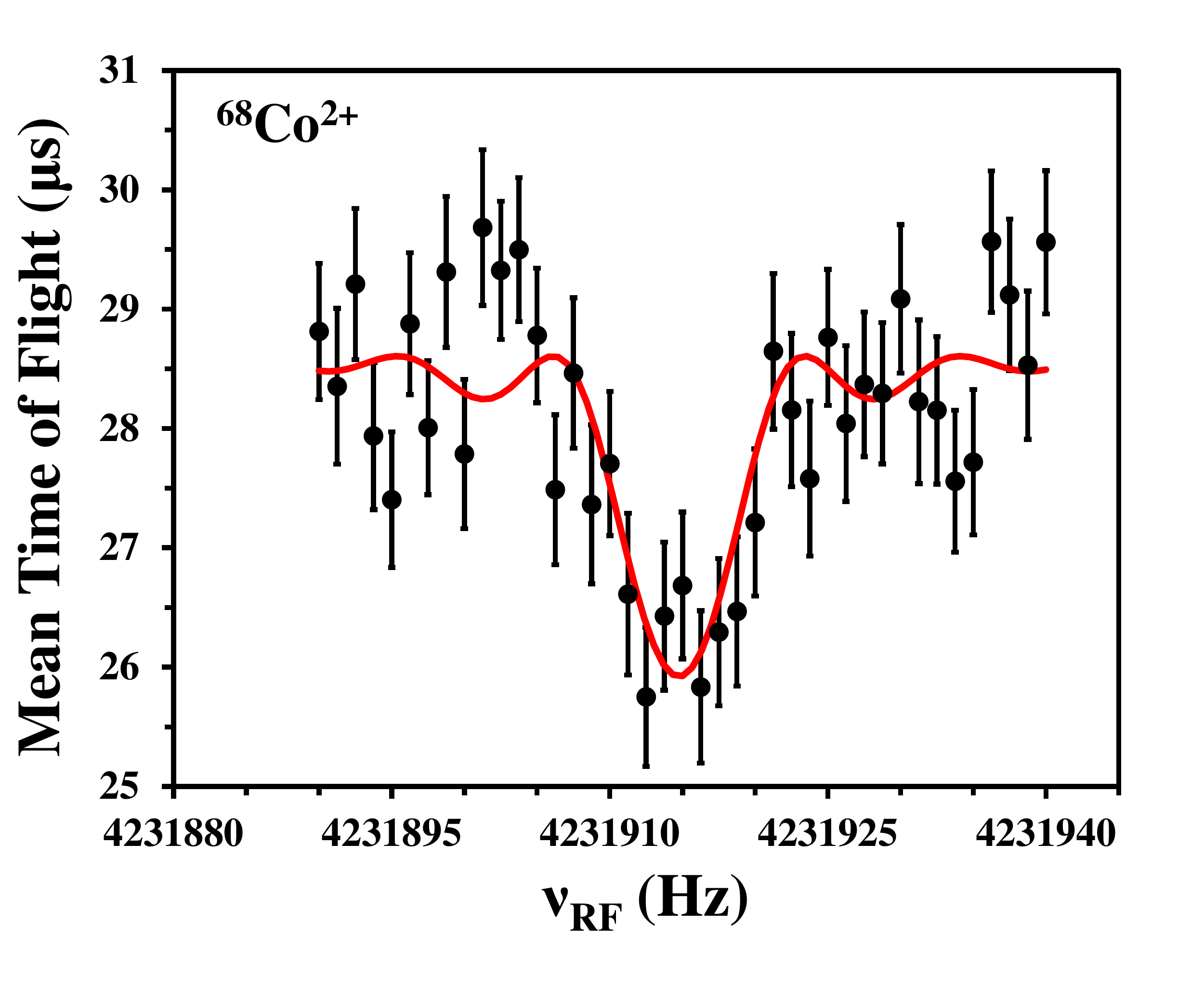}
\caption{\label{fig2}Example of the \nuc{68}{Co}$^{2+}$ TOF-ICR resonances used for the determination of $\nu_c$(\nuc{68}{Co}$^{2+}$).  This resonance contains 6674 ions.  The solid line is a theoretical line shape \cite{Konig1995} fit to the data.}
\end{figure}

The cyclotron frequency is then determined using the time-of-flight ion cyclotron resonance (TOF-ICR) technique \cite{Konig1995}.  In this method, the applied RF frequency is scanned near the expected cyclotron frequency and the ions' time-of-flight is measured from the Penning trap to an MCP detector outside the 9.4-T magnet.  At the point where the applied RF is in resonance with the ions' cyclotron frequency, a reduced time-of-flight is observed.  A typical TOF-ICR resonance curve obtained in this experiment is presented in Fig.~\ref{fig2}.  Cyclotron frequency measurements of $^{68,69}$Co$^{2+}$ ions were alternated with measurements of stable reference ions with well-known masses in order to monitor the magnetic field strength.  By monitoring the magnetic field strength and measuring the cyclotron frequency, Eq. (3) could be used to precisely determine the ion masses.

\begin{table*}[t]
\caption{\label{TableI} Measured frequency ratios, $\nu_{c}$/$\nu_{c,\text{ref}}$, calculated atomic mass and mass excess (ME) values, and their comparison to the values from 2016 Atomic Mass Evaluation.  Differences in the mass excess values, $\Delta$ME = $\text{ME}_{\text{LEBIT}} - \text{ME}_{\text{AME2016}}$, are also listed.}
\begin{ruledtabular}
\begin{tabular}{ccccccc}
\multicolumn{1}{c}{Ion} & \multicolumn{1}{c}{Reference} & \multicolumn{1}{c}{Frequency Ratio} & \multicolumn{1}{c}{Mass (u)} & \multicolumn{1}{c}{ME (keV)}  & \multicolumn{1}{c}{AME2016 (keV)}  & \multicolumn{1}{c}{$\Delta$ME (keV)} \\
\hline
\multirow{2}*{$^{68}$Co$^{2+}$} & $^{16}$O$^{18}$O$^+$ & $1.000\ 641\ 552(70)$ & $67.944\ 559\ 2(48)$ & $-51\ 642.8(4.4)$  &  \multirow{2}*{$-51\ 930(190)$} & 290(190) \\
   & $^{34}$S$^+$ & $0.999\ 870\ 11(12)$ & $67.944\ 559\ 3(82)$ & $-51\ 642.6(7.6)$  &  & 290(190) \\
\hline
{$^{69}$Co$^{2+}$} & $^{39}$K$^+$ &  $1.130\ 267\ 90(24)$ & $68.946\ 093(15)$ & $-50\ 214(14)$  &  {$-50\ 280(140)$} & 66(140) \\
\end{tabular}
\end{ruledtabular}
\end{table*}

\section{Results and Discussion}

Each Co measurement was preceded and followed by a reference mass measurement.  \nuc{68}{Co}$^{2+}$ measurements were performed on two separate occasions approximately one year apart; on the first occasion \nuc{16}{O}\nuc{18}{O}$^{+}$ was used as a reference and on the second occasion \nuc{34}{S}$^+$ was used as a reference.  \nuc{39}{K}$^+$ was used as a reference for \nuc{69}{Co}$^{2+}$.  The reference measurement frequencies were then linearly interpolated to determine the reference frequency at the time the ion of interest was measured.  The atomic mass of the ion of interest is then calculated from the ratio of the cyclotron frequencies of the two species by

\begin{equation}
m = \left[m_{\textnormal{ref}} - \frac{q_{\textnormal{ref}}}{e}\cdot m_e\right] \times \frac{q}{q_{\textnormal{ref}}}\frac{1}{r} + 2m_e
\end{equation}
where $r$ is the ratio $\frac{\nu_c}{\nu_{c,\textnormal{ref}}}$ and $e$ is the elementary charge.  The ionization potentials of all species and the molecular binding of \nuc{16}{O}\nuc{18}{O}$^{+}$ are not included in these calculations as they are all $<$ 20~eV and do not contribute at our level of uncertainty \cite{NIST_ASD}.  On the first occasion \nuc{68}{Co} was measured, seven frequency ratios containing a total of 4779 \nuc{68}{Co}$^{2+}$ ions were recorded with a weighted average $r_1=1.000641552(70)$.  A near-unity Birge ratio \cite{Birge1932} of 0.93(18) indicates that additional statistical effects are unlikely.  On the second occasion \nuc{68}{Co} was measured, ten frequency ratios containing a total of 23129 \nuc{68}{Co}$^{2+}$ ions were recorded with a weighted average $r_2=0.99987011(12)$.  On this occasion the Birge ratio was 1.96(15), so the statistical uncertainty was inflated by multiplication with the Birge ratio.  This commonly-used Birge adjustment is valid in the case where all uncertainties are underrated by the same common factor, which is taken to be true for measurements of equal reliability.  Five frequency ratios were recorded for \nuc{69}{Co}, with a weighted average $r=1.13026790(24)$ and a Birge ratio of 1.22(21).  This statistical uncertainty has also been inflated by multiplication with the Birge ratio.  Systematic effects such as minor trap misalignment in the magnetic field and deviations from a purely quadrupole electric potential result in small frequency shifts dependent on the mass-to-charge difference from the reference ion.  These shifts have previously been evaluated at LEBIT and contribute only at a level of $2.0 \times 10^{-10} /(q/u)$ \cite{Gulyuz2015}, which is negligible for all frequency ratios considered here.  Taking a weighted average of the results from the two independent measurements of \nuc{68}{Co} yields a mass excess of $\textnormal{ME}(^{68}\textnormal{Co}) = -51642.7(3.8) \unit{keV}$, and the mass excess of \nuc{69}{Co} was determined to be $\textnormal{ME}(^{69}\textnormal{Co}) = -50214(14) \unit{keV}$.

A summary of these results from LEBIT and comparison with the 2016 Atomic Mass Evaluation (\textsc{Ame}2016) \cite{Wang2016} is presented in Table~I.  Before drawing conclusions about these results, however, we must consider the possible presence of isomers.  Beta-decay experiments have reported the existence of two beta-decaying states in \nuc{68}{Co} \cite{Mueller2000} and recently a second beta-decaying state in \nuc{69}{Co} as well \cite{Liddick2015}.  In the case of \nuc{68}{Co}, the \textsc{Ame}2016 value comes from time-of-flight measurements which would not resolve an isomeric state from the ground state if both were present.  However, any isomers present in these measurements would pull the \textsc{Ame}2016 ground state mass even further away from the LEBIT results.  In the past, LEBIT has on several occasions demonstrated the ability to resolve ground and isomeric states when the states are populated at roughly equal proportions in the fragmentation process \cite{Block2008, Ferrer2010, Valverde2015}.  This is marked by a signature double peak in the TOF-ICR resonance, as seen in Fig.~2 of Ref. \cite{Valverde2015}.  Since only a single peak in the TOF-ICR resonance was observed for both \nuc{68}{Co} and \nuc{69}{Co}, an additional step was implemented to examine which state was present in the rare isotope beam delivered to LEBIT.

\begin{figure}
\includegraphics[scale=0.35]{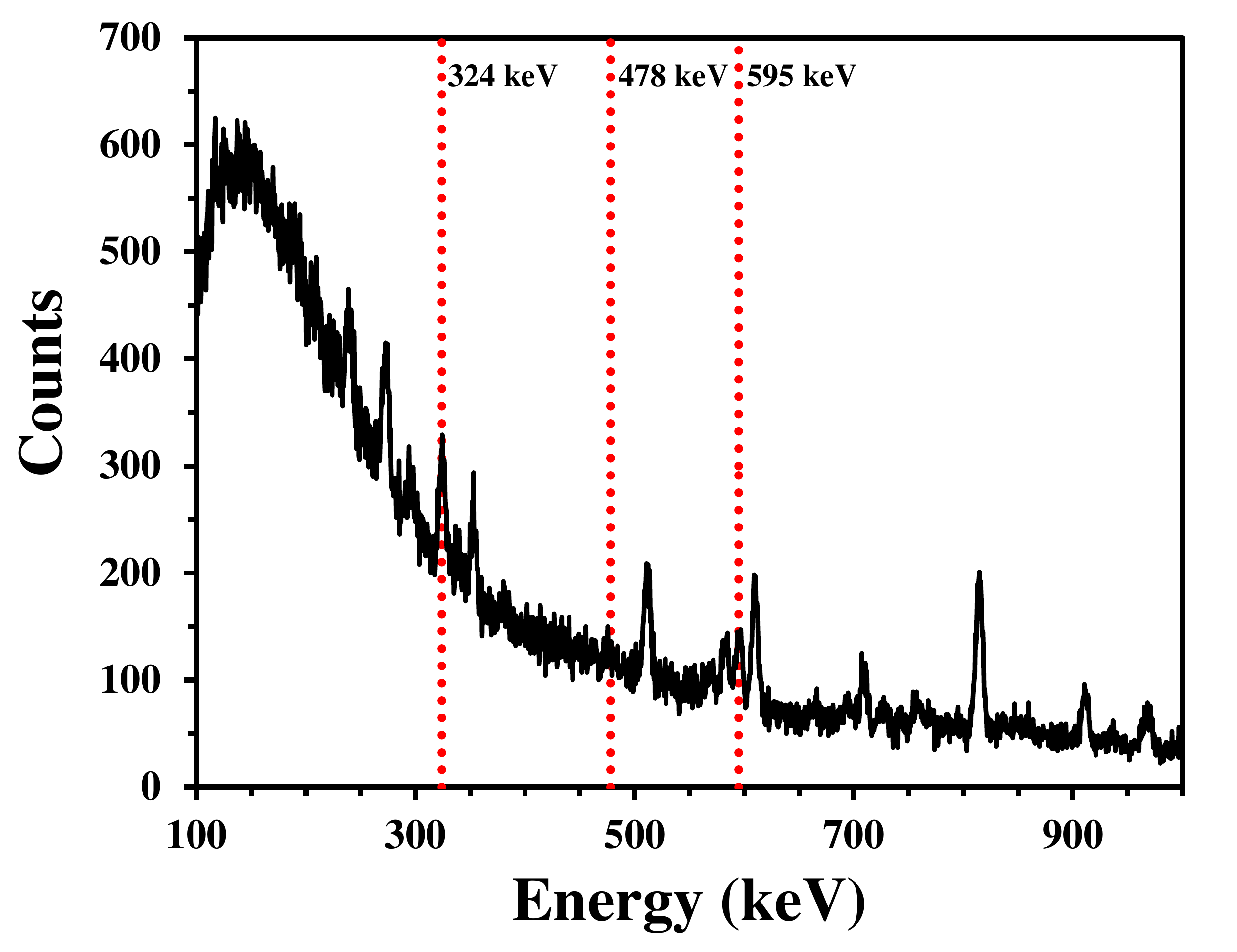}
\caption{\label{fig3}The beta-delayed gamma-ray spectrum detected following the decay of $^{68}\textnormal{Co}$.  Dashed vertical lines mark the energies of gamma rays known to follow primarily decay from only one of the two beta-decaying states of $^{68}\textnormal{Co}$.  Gamma rays at 478~keV follow decay from the low-spin state and gamma rays at 324~keV and 595~keV follow decay from the high-spin state.  The presence of clear peaks at 324~keV and 595~keV and the lack of a peak at 478~keV suggests that the high-spin beta-decaying state of $^{68}\textnormal{Co}$ was measured in this experiment.  Other peaks in the spectrum have all been identified from either natural background radiation or $^{68}\textnormal{Co}$ decays common to both beta-decaying states.}
\end{figure}

Prior to delivery of \nuc{68}{Co}$^{2+}$ ions from the gas cell to LEBIT, they were collected in front of a high purity germanium detector.  As the Co decayed, beta-delayed gamma rays were detected with the Ge detector, and the resulting spectrum was compared with the current literature available for \nuc{68}{Co} beta decay \cite{Mueller2000, Flavigny2015, Prokop2016}.  While there is disagreement in the literature regarding the relative gamma intensities, there is agreement that decay from the low-spin state produces a 478~keV gamma ray and any production of 595~keV or 324~keV gamma rays are at very low intensities ($< 1\%$).  Decay from the high-spin state of \nuc{68}{Co}, on the other hand, is associated with higher intensity production of the 595~keV and 324~keV gammas ($\sim 32\%$ and $\sim 38\%$, respectively) and no production of the 478~keV gamma ray.  As shown in Fig.~\ref{fig3}, the gamma spectrum collected as part of this work shows clear peaks at 595~keV and 324~keV, while no evidence of a peak at 478~keV is present.  This demonstrates the presence of the high-spin beta-decaying state in \nuc{68}{Co}, and does not support the presence of any of the low-spin beta-decaying state.

It is worth noting that the absolute intensity of the 478~keV gamma in the low-spin state decay ($\sim 6\%$) is substantially lower than the reported intensities of the 595~keV and 324~keV gammas in the high-spin state decay ($\sim 32\%$ and $\sim 38\%$, respectively).  Additionally, the high-spin state has a shorter half-life than the low-spin state (0.23(3)~s as compared to 1.6(3)~s) \cite{Mueller2000}.  Accounting for the size of the peaks in the gamma spectrum, the detector efficiency at each energy, the relative intensities of each peak, and the expected decay losses from the gamma detector to the Penning trap, it was calculated that the low-spin state could have been present at a level just below detectability at the Ge detector and still been the dominating beam component at the Penning trap.  However, given that there is no positive evidence of a second state in the gamma spectrum or in the TOF-ICR resonance and the gamma spectrum shows compelling evidence for the presence of the high-spin state, we assume that this was in fact the state measured in the Penning trap.

\nuc{69}{Co}$^{2+}$ was delivered to LEBIT at a very low rate ($<$ 0.5~ions/second), so it was not possible to obtain a gamma spectrum as was done with \nuc{68}{Co}.  However, the (1/2$^-$) beta-decaying state proposed by Liddick \textit{et al.} \cite{Liddick2015} was only observed in \nuc{69}{Co} when produced by beta decay from \nuc{69}{Fe}.  In that same experiment, \nuc{69}{Co} was also produced directly by projectile fragmentation, and there only the shorter-lived 7/2$^-$ state was observed.  As \nuc{69}{Co} was only produced directly by projectile fragmentation in this work (albeit with a different primary beam) and only one state was observed here as well, we assume that the \nuc{69}{Co} mass determination reported here corresponds to the 7/2$^-$ state.

With these assumptions, we conclude that, in both \nuc{68}{Co} and \nuc{69}{Co} the one state measured in the Penning trap was the beta-decaying state with the higher spin.  However, in both cases, the ordering of the beta-decaying states is still unknown.  Mueller \textit{et al.} proposed spin and parity assignments for the two states in \nuc{68}{Co} based on angular momentum coupling of the two beta-decaying configurations in \nuc{69}{Ni} with an $f_{7/2}$ proton hole \cite{Mueller2000}, which would suggest a high-spin (7$^-$) ground state and a low-spin (3$^+$) isomer.  This argument follows the rules for angular momentum coupling of particle-hole configurations in odd-odd nuclei laid out by Brennan and Bernstein \cite{Brennan1960}.  While others have suggested alternate spin and parity assignments for the low-spin state \cite{Liddick2012, Flavigny2015}, none have yet offered any contradiction to this ordering.

In the case of \nuc{69}{Co}, one of the beta-decaying states is believed to be 7/2$^-$ based on the $\pi f_{7/2}^{-1}$ configuration observed in all other odd-$A$ Co isotopes, and the other beta-decaying state proposed by Liddick \textit{et al.} is described as a (1/2$^-$) prolate-deformed intruder state attributed to proton excitations across the Z=28 shell as has been suggested for \nuc{65}{Co} and \nuc{67}{Co} \cite{Pauwels2008}.  The (1/2$^-$) state approaches the 7/2$^-$ ground state near $N$=40 and becomes isomeric, possibly even crossing the ground state, at \nuc{69}{Co}.  No comment on the ordering of these two states can be made in \cite{Liddick2015}, but their separation is limited to $<$467~keV or $<$661~keV depending on the assumed strength of the unobserved M3 $\gamma$-ray transition.

In order to shed light on the ordering of the beta-decaying states in \nuc{68}{Co} and \nuc{69}{Co}, we performed \textit{ab initio} calculations using the valence-space in-medium similarity renormalization group (VS-IMSRG) \cite{Tsuk12SM,Bogn14SM,Stro16TNO,Herg16PR,Stroberg2017} framework based on two-nucleon (NN) and three-nucleon (3N) forces from chiral effective field theory \cite{Epel09RMP,Mach11PR}. In particular we use the SRG-evolved \cite{Bogn10PPNP} 1.8/2.0 (EM) interaction from Refs.~\cite{Hebe11fits,Simo16unc}, which predicts realistic saturation properties of infinite matter and has been shown to reproduce well ground-state and separation energies from the $p$-shell to the tin isotopes \cite{Simo17SatFinNuc,Morr17sn100}. We then use the Magnus formulation of the IMSRG \cite{Morr15Magnus} to construct an approximate unitary transformation to decouple a given valence space Hamiltonian (and core energy) to be diagonalized with a standard shell-model code \cite{Brow14NuShellX}. 

We consider two different valence space strategies for cobalt isotopes. The first uses standard $0\hbar\omega$ spaces for both protons and neutrons:  for protons, the $pf$ shell; for neutrons, the $pf$ shell above a \nuc{40}{Ca} core for $N<40$, and the $sdg$ shell above a \nuc{60}{Ca} core for $N>40$. These spaces allow for no neutron excitations at $N=40$, so we know \emph{a priori} that calculations in this vicinity will be unreliable. Since this is also the region of interest for the current measurements, we also decouple a cross-shell $p_{1/2}f_{5/2}g_{9/2}$ neutron space using a $^{52}$Ca core. 

With the $p_{1/2}f_{5/2}g_{9/2}$ space, the ground state in \nuc{68}{Co} was found to be a 2$^-$, which agrees with the spin-parity of the low-spin beta-decaying state suggested by Flavigny \textit{et al.} \cite{Flavigny2015}.  However, the calculations predict a large number of states below 1~MeV, so a definitive prediction of the ground-state is not possible given present theoretical uncertainties. In the case of \nuc{69}{Co}, the ground state was calculated to be 7/2$^-$, consistent with the surrounding odd-mass Co isotopes.

It is clear that additional work is needed to clarify the orderings of the two beta-decaying states in \nuc{68}{Co}. Mueller \textit{et al.} suggest a high-spin ground state \cite{Mueller2000} while the VS-IMSRG calculations suggest a low-spin ground state.  However, with the large number of close-lying states, neither proposal is presented with a high degree of confidence.  For the purpose of examining mass surfaces in the regions of \nuc{68}{Ni}, the high-spin state of \nuc{68}{Co} measured in this work is treated as the ground state.  Should future work challenge this assignment, it is worth noting that the results presented in this work will still be valuable for a precise determination of the ground-state mass if the excitation energy of the isomeric state has been measured.

\begin{figure}
\includegraphics[scale=0.35]{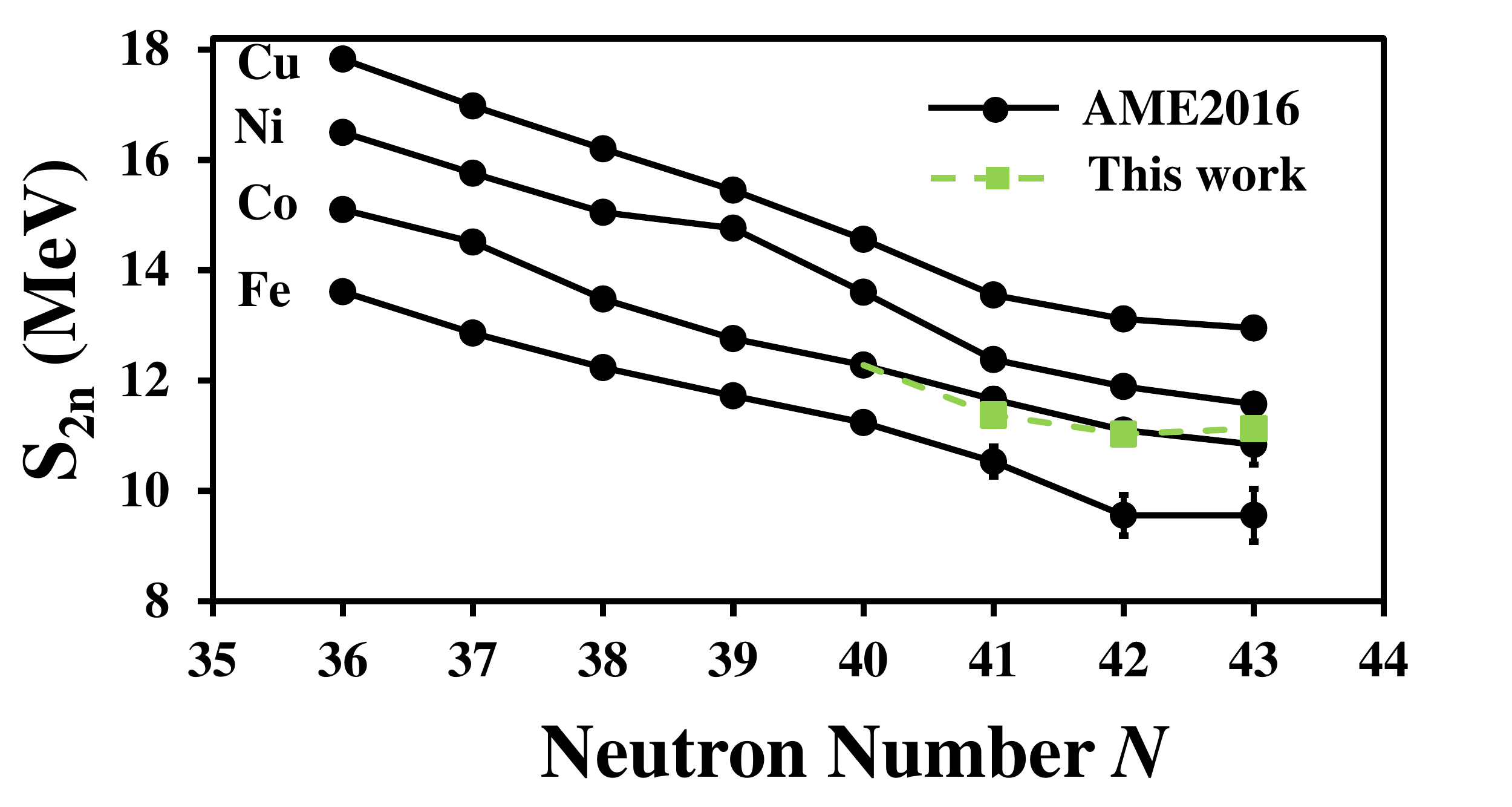}
\caption{\label{fig4}(color online) Two-neutron separation energies $S_{2n}$ plotted as a function of neutron number $N$ for isotopes of Fe, Co, Ni, and Cu.  Lighter (green) square points correspond to new Co values from this work and darker (black) circles correspond to data from \textsc{Ame}2016 \cite{Wang2016}.}
\end{figure}

\begin{figure}
\includegraphics[scale=0.35]{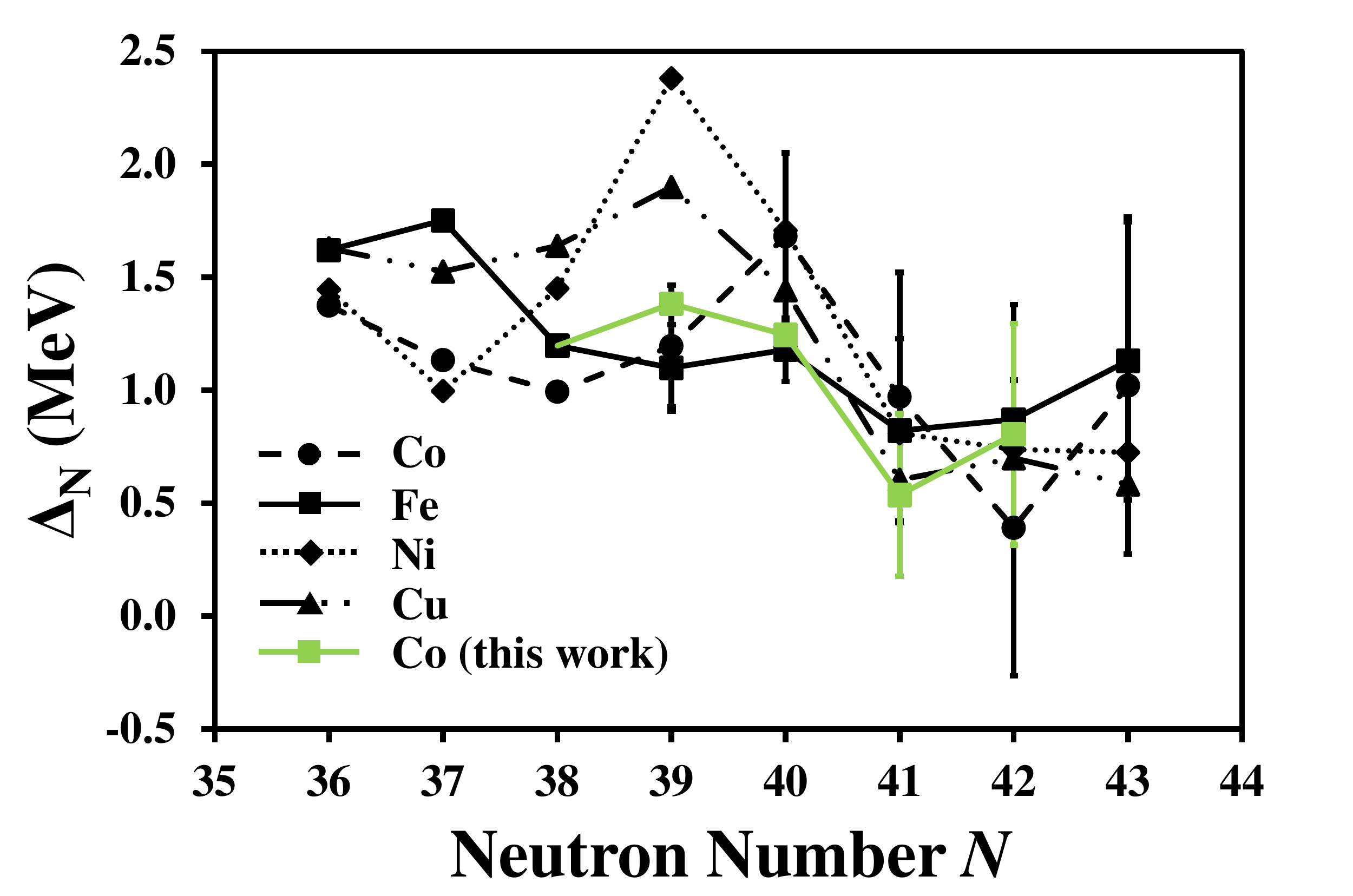}
\caption{\label{fig5}(color online) Neutron shell gap $\Delta_{N}$ plotted as a function of neutron number $N$ for isotopes of Fe, Co, Ni, and Cu.  The lighter (green) curve corresponds to new Co values from this work and the darker (black) curves correspond to data from \textsc{Ame}2016 \cite{Wang2016}.}
\end{figure}

The trends in $S_{2n}$ are presented in Fig.~\ref{fig4}, showing the \textsc{Ame}2016 data and results from this work.  While the \textsc{Ame}2016 data shows a fairly linear trend along the Co chain from $N$=39 to $N$=42, the new LEBIT data demonstrates a substantial (287~keV) reduction in binding at $N$=41, creating a small kink in the $S_{2n}$ chain which might suggest a minor subshell closure at $N=40$.  To examine this more quantitatively, we have also calculated the neutron shell gap parameter, defined as
\begin{equation}
\Delta_{N}(N,Z) = S_{2n}(N,Z) - S_{2n}(N+2,Z).
\end{equation}
$\Delta_N$ is plotted as a function of neutron number for Fe, Co, Ni, and Cu in Fig.~\ref{fig5} in the region of $N=40$.  While the new measurement of \nuc{68}{Co} significantly reduces $\Delta_N$ at $N$=41 in the Co chain, it also increases $\Delta_N$ at $N=39$ such that no relative peak is observed at $N=40$.  However, the high precision of the LEBIT results reveals a small but significant enhancement of $\Delta_N$ at $N$=39, somewhat smaller than what can already be seen at $N$=39 in the Cu chain and significantly smaller than the peak at $N$=39 seen in the Ni chain.  While it does not make any sense from a shell model perspective to consider this a sign of a shell or subshell closure, this does seem to suggest some as yet unexplained behavior resulting in additional stability at $N$=39 in this region.  

\begin{figure}
\includegraphics[scale=0.35]{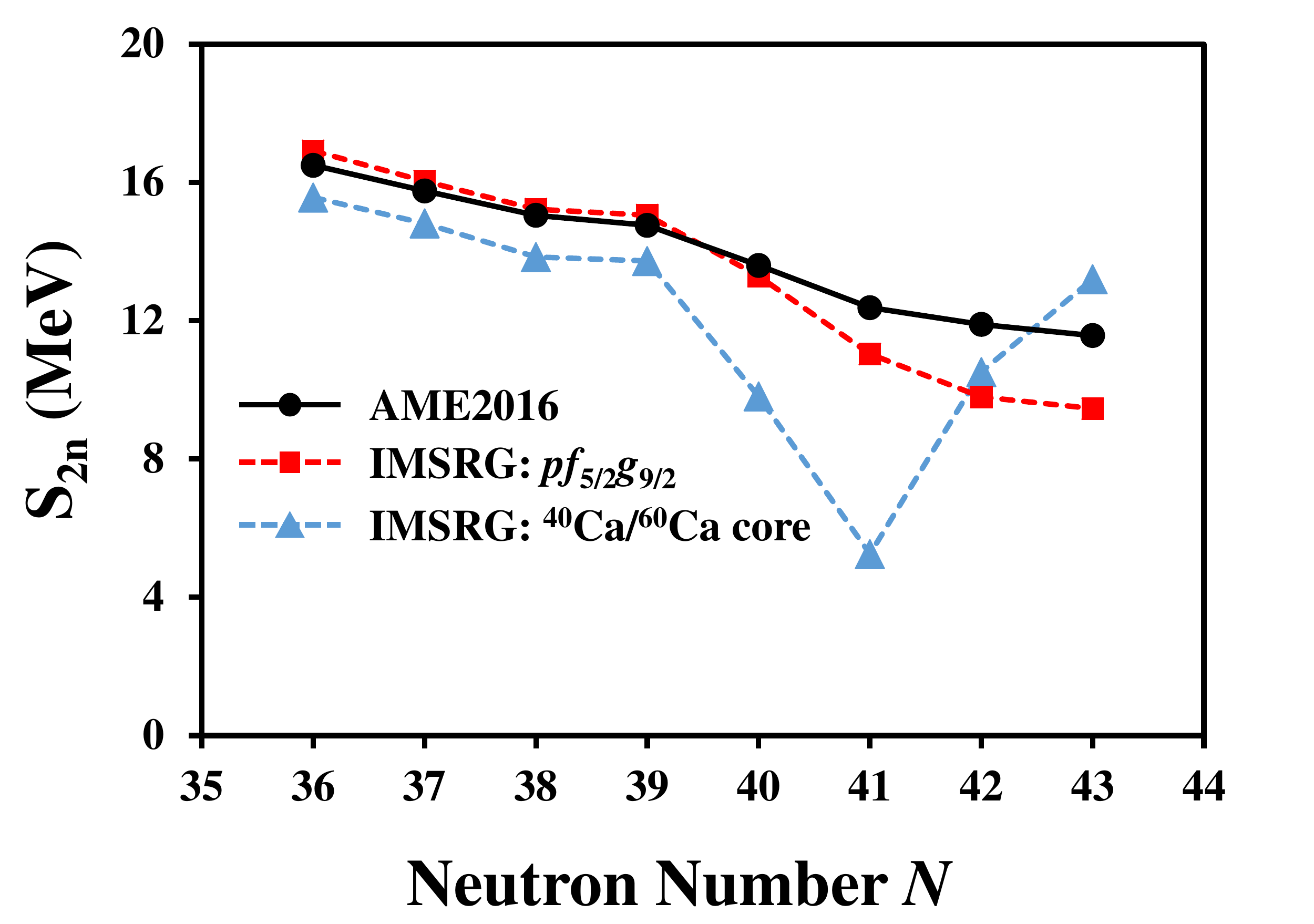}
\caption{\label{fig6}(color online) Two-neutron separation energies $S_{2n}$ plotted as a function of neutron number $N$ for the Ni isotopic chain.  The solid black data corresponds to data from \textsc{Ame}2016, and the dashed lines correspond to IM-SRG theoretical calculations.  Blue triangles use a \nuc{40}{Ca} core for $N \leq$ 40 (\textit{pf} neutron valence space) and a \nuc{60}{Ca} core for $N>$ 40 (\textit{sdg} neutron valcence space).  Red squares use a $pf_{5/2}g_{9/2}$ valence space for protons and neutrons for the entire chain starting from a \nuc{56}{Ni} core.}
\end{figure}

Interestingly, the IMSRG calculations discussed previously capture this surprisingly large binding at $N=39$ in the Ni chain.  A comparison of the IMSRG calculations and the \textsc{Ame}2016 data for the Ni chain in this region is presented in Fig.~\ref{fig6}.  IMSRG calculations performed within a single harmonic oscillator shell (using only $pf$ shell for $N \leq$ 40 and $sdg$ shell for $N >$ 40) demonstrate significant underbinding and result in an unphysical discontinuity crossing $N=40$. As mentioned above, this discontinuity is well understood to be due to the lack of allowed neutron excitations near $N=40$, which is an artifact of this particular choice of valence space; indeed the binding is significantly improved for $N <$ 40, and the discontinuity across $N=40$ disappears when a $pf_{5/2}g_{9/2}$ neutron space is used instead. Both cases capture the unexpectedly large binding at $N=39$ observed in the \textsc{Ame}2016 data.

\section{Summary}

The first Penning trap mass measurements of \nuc{68}{Co} and \nuc{69}{Co} have been completed at the National Superconducting Cyclotron Laboratory.  These measurements reduce the atomic mass uncertainties by more than an order of magnitude from the \textsc{Ame}2016, allowing for detailed studies of nuclear structure in the area near $Z$=28, $N$=40.  Although further studies are needed to definitively establish the ordering of the two beta-decaying states in \nuc{68,69}{Co}, no evidence for a substantial subshell closure across $N$=40 was observed in the $_{27}$Co isotopes, consistent with $S_{2n}$ studies already completed for $Z \geq$ 28.  We have also found that the unexpectedly high ground state binding energy at $N$=39 in the Ni isotopes is well-reproduced by IMSRG structure calculations, and the experimental evidence presented here suggests that this behavior may persist at a lesser extent in the neighboring Cu and Co isotopic chains.

\begin{acknowledgments}
 This work was conducted with the support of Michigan State University and the National Science Foundation under Grant No. PHY-1102511, No. PHY-1419765, and No. PHY-1713857. The work leading to this publication has also been supported by a DAAD P.R.I.M.E. fellowship with funding from the German Federal Ministry of Education and Research and the People Programme (Marie Curie Actions) of the European Union's Seventh Framework Programme (FP7/2007/2013) under REA grant agreement no. 605728. This work was supported in part by the Natural Sciences and Engineering Research Council of Canada (NSERC). We thank K. Hebeler, J. Simonis and A. Schwenk for providing the 3N matrix elements used in this work and for valuable discussions. Computations were performed with an allocation of computing resources at the J\"ulich Supercomputing Center (JURECA).
\end{acknowledgments}

\end{document}